\documentclass[]{llncs}
\usepackage[colorinlistoftodos,prependcaption,textsize=tiny]{todonotes}
\usepackage{graphicx,subfigure}
\usepackage{todonotes}
\usepackage[cmex10]{amsmath}
\usepackage{algpseudocode,algorithm}
\usepackage{url}
\usepackage{amssymb}
\usepackage{listings}
\usepackage{url}
\usepackage{fixme}
\fxusetheme{color}
\fxsetup{
     status=draft,
     author=,
     layout=inline, 
     theme=color
}

\newcommand{\I}[1]{\textit{#1}}

\newcommand{\upup}{\vspace*{-0.3em}}

\begin{document}

\definecolor{darkblue}{rgb}{0,0,.6}
\definecolor{darkred}{rgb}{.6,0,0}
\definecolor{darkgreen}{rgb}{0,.6,0}
\definecolor{red}{rgb}{.98,0,0}
\definecolor{gray}{rgb}{.6,.6,.6}

\title{Closing the Performance Gap with Modern C++}

\author{Thomas Heller\inst{1,5} \and Hartmut Kaiser\inst{2,5}  \and Patrick Diehl\inst{3,5} \and Dietmar Fey\inst{1} \and Marc Alexander Schweitzer\inst{3,4}}

\institute{Computer Science 3, Computer Architectures, Friedrich-Alexander-University \and Center for Computation and Technology, Louisiana State University \and Institute for Numerical Simulation, University of Bonn \and Meshfree Multiscale Methods, Fraunhofer SCAI, Schloss Birlinghoven \and The STEllAR Group (\url{http://stellar-group.org})}

\maketitle

\begin{abstract}

On the way to Exascale, programmers face the increasing challenge of having to support multiple
hardware architectures from the same code base. At the same time, portability of code and
performance are increasingly difficult to achieve as hardware architectures are becoming more
and more diverse. Today's heterogeneous systems often include two or more completely distinct
and incompatible hardware execution models, such as GPGPU's, SIMD vector units, and
general purpose cores which conventionally have to be programmed using separate tool
chains representing non-overlapping programming models. The recent revival of interest
in the industry and the wider community for the C++ language has spurred a remarkable
amount of standardization proposals and technical specifications in the arena of concurrency
and parallelism. This recently includes an increasing amount of discussion around the need
for a uniform, higher-level abstraction and programming model for parallelism in the C++
standard targeting heterogeneous and distributed computing. Such an abstraction should
perfectly blend with existing, already standardized language and library features, but
should also be generic enough to support future hardware developments. In this paper, we
present the results from developing such a higher-level programming abstraction for
parallelism in C++ which aims at enabling code and performance portability over a wide
range of architectures and for various types of parallelism. We present and compare
performance data obtained from running the well-known STREAM benchmark ported to
our higher level C++ abstraction with the corresponding results from running it natively.
We show that our abstractions enable performance at least as good as the comparable
base-line benchmarks while providing a uniform programming API on all compared
target architectures.
\end{abstract}


\section{Introduction}
\upup

The massive local parallelism available on today's and tomorrow's systems poses one of the
biggest challenges to programmers, especially on heterogeneous architectures, where
conventional techniques require to develop and tune independent code bases for each of
the separate parts of the machine. This paper focuses on how to address portability in terms
of code and performance when developing applications targeting heterogeneous
systems. More and more systems come on-line which consist of more than one
hardware architecture, all of it made available to the developers through often
independent and orthogonal tool-chains.

With the recently growing interest in the community in C++ and the increased activity
towards making all of the available parallelism of the machine available through native
C++ language features and library facilities, we see an increasing necessity in developing
higher level C++ APIs which ensure a high level of portability of code while providing the
best possible performance. At the same time, such APIs have to provide a sufficient amount of
generality and flexibility to provide a solid foundation for a wide variety of application
use cases. GPGPU vendors have started to make their C++ tool chains more conforming
with the newest C++11/C++14 Standards~\cite{cpp_standard}, as demonstrated for instance by recent
versions of NVidia's CUDA~\cite{cuda} or the newer HCC compiler~\cite{hcc} as provided by AMD. Unfortunately,
there are no usable standards-conforming library solution available yet which would help
in writing C++ code which is portable across heterogeneous architectures.

One of the key problems to solve while developing such higher level library abstractions
is to provide facilities to control and coordinate the placement of data in conjunction with
the location of the execution of the work on this data. We describe the result of our research
in this direction, provide a proof of concept optimization, and present performance results
gathered from comparing native implementations of the STREAM benchmark for OpenMP~\cite{McCalpin1995}
and CUDA~\cite{Deakin2015} with an equivalent application written based on our design. We show that
there is essentially no performance difference between the original benchmarks and
our results.

Our presented implementation of C++ algorithms is fully conforming to the specification
to be published as part of the C++17 Standard~\cite{parallelism_ts}. It is based on HPX~\cite{hpx_v0.9.99}, a parallel runtime system
for applications of any scale. For our comparisons with the native
OpenMP and CUDA benchmarks we use the same sources demonstrating a high degree
of portability of code and performance. The used parallel algorithms are conforming to the
latest C++17 Standard and are designed to be
generic, extensible and composable.

In the remaining part of this paper we describe related work (Sec.~\ref{sec:related_work}),
talk about locality of data and work (Sec.~\ref{sec:concepts}), describe our implementations
(Sec.~\ref{sec:implementations}), show the results (Sec.~\ref{sec:results}), and summarize
our findings (Sec.~\ref{sec:conclusions}).

%
%
%
%
%
%
%

\section{Related Work}
\label{sec:related_work}
\upup

The existing solutions for programming accelerators mostly have in common that they are based
either on OpenCL~\cite{1997openmp} or on CUDA~\cite{cuda} as their backends. Table~\ref{tab::overview} shows an
overview of the different approaches existing today. The most prominent in that
regard are pragma based language extensions such as OpenMP and
OpenACC~\cite{openacc}. The pragma solutions naturally don't offer good support
for C++ abstractions. In order to get better C++ language integration, software has to directly
rely on newer toolchains directly supporting C++, such as recent versions of CUDA, the
newer HCC compiler~\cite{hcc}, or SYCL~\cite{sycl}.

Generic higher level abstractions are also provided by various library based
solutions such as Kokkos~\cite{carteredwards20143202}, raja~\cite{raja},
Thrust~\cite{hoberock2010thrust}, and Bolt~\cite{bolt}. Those attempt to offer
higher level interfaces similar but not conforming to the parallel algorithms specified
in the upcoming C++17 -Standard~\cite{cpp17_standard}. One of the contributions
of this paper is to provide standards-conforming implementations of those parallel algorithms
combined with truly heterogeneous solutions enabling transparent locality
control, a feature not available from the listed existing libraries.

In addition, we aim to provide a solution for all existing accelerator architectures, that is
not limited to either OpenCL or CUDA based products providing a modern C++
programming interface.

\begin{table}
\centering
\begin{tabular}{llll}
\hline
name & type & hardware support & \\
\hline
OpenMP & pragmas & cpu, accelerators & \cite{1997openmp} \\
OpenACC  & pragmas & accelerators & \cite{openacc} \\
HCC & compiler & OpenCL, HSA  & \cite{hcc}\\
CUDA & compiler & CUDA & \cite{cuda}\\
SYCL & compiler & OpenCL & \cite{sycl} \\
Kokkos & library & OpenMP, CUDA & \cite{carteredwards20143202} \\
raja & library & OpenMP, CUDA & \cite{raja} \\
thrust & library & CUDA, TBB, OpenMP & \cite{hoberock2010thrust} \\
bolt  & library & C++Amp, OpenCL, CPU & \cite{bolt} \\
\hline \\
\end{tabular}
    \caption{overview of different approaches: pragma based solutions, low level compiler and libraries to leverage different architectures.}
    \label{tab::overview}
\end{table}

\section{Locality of Work and Data}
\label{sec:concepts}
\upup

\begin{figure}[ht]
    \begin{center}
    \includegraphics[width=0.8\linewidth]{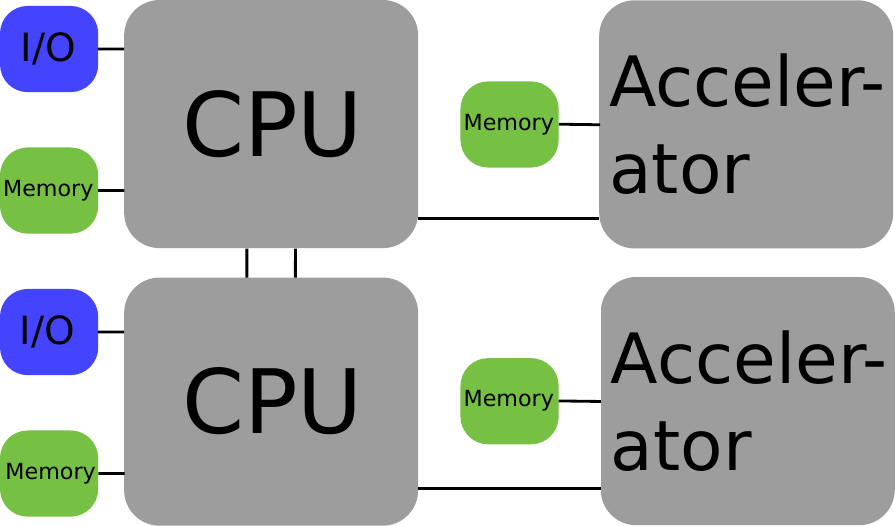}
    \end{center}
\upup
\upup
    \caption{This Figure shows an example of a typical heterogeneous
    architecture that is composed of multiple CPUs containing different physical
    blocks of memory as well as Accelerators and Network Interfaces with their own
    discrete memory locations, which are connected through a common bus.\upup}
    \label{fig:architecture}
\end{figure}

Modern computing architectures are composed of various different levels of
processing units and memory locations. Fig.~\ref{fig:architecture} shows an example
for such architectures that are a common in today's nodes for GPU accelerated
supercomputers. Tomorrow's systems will be composed of even more complex memory
architectures. In addition, when for instance looking at autonomous driving applications
requiring a huge amount of processing power, the diversity of different processing units
as well as different memory locations will increase.

In order to program these architectures efficiently it is important to
place the data as close as possible to the site where the execution has to take
place. As such, we need APIs that are able to effectively and transparently express
the data placement on and data movement to concrete memory locations (or
places) in a system. We also need APIs allowing to coordinate the desired data
placement with fine control over defining the execution site from where the code
will access the data.

\begin{figure}[ht]
	\centering
    \includegraphics[width=0.5\linewidth]{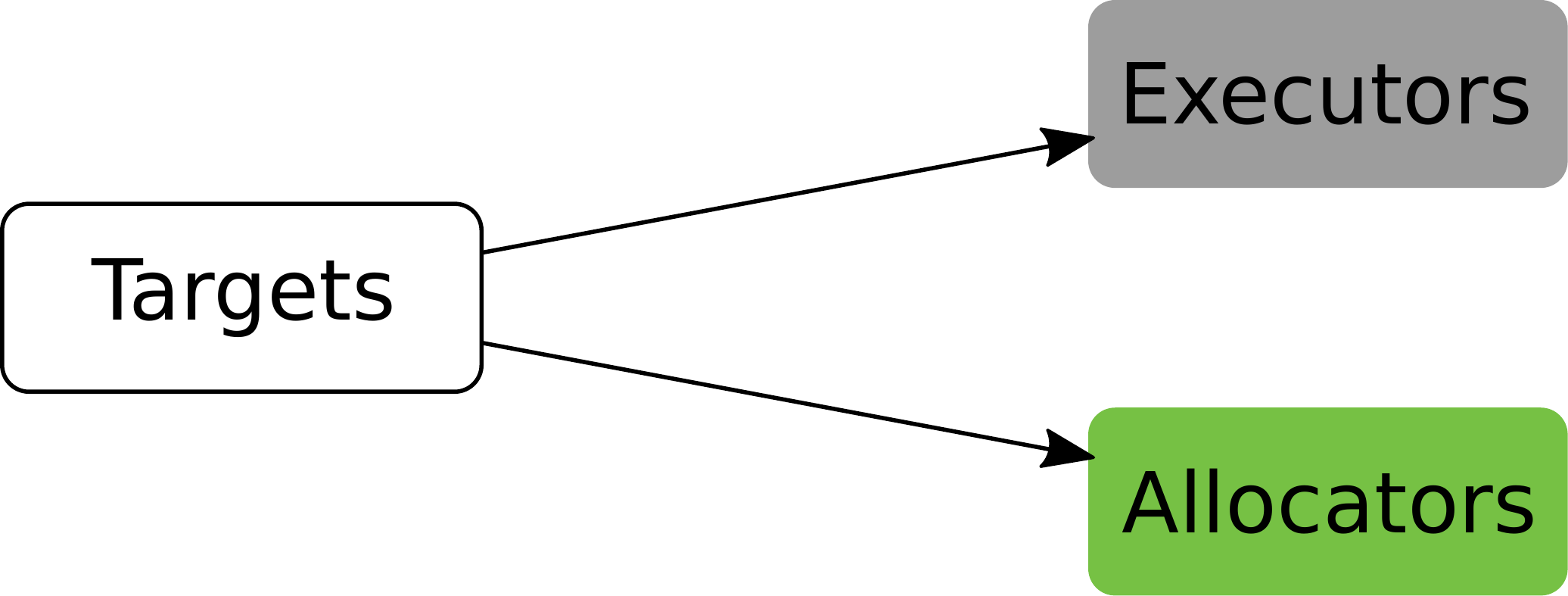}
\upup
\upup
    \caption{This figure shows the relation between targets, memory allocation and
    work execution; A target is the link between co-locating memory allocation and
    execution of tasks close to the memory location. It is used to transparently
    express the notion of a `place' in the system to both, allocation and execution.}
    \label{fig:target_relation}
\end{figure}

This paper proposes concepts and APIs that are rooted within the C++ language
and Standard Library to create an expressive, performant, and extensible
way to control locality of work and data by refining the \I{allocator} concept already defined
in the C++ standard as well as using the proposed \I{executor} concept. These are tied
together by defining \I{targets}, which represent places in a system, to properly
co-locate placement of data and execution of work (see Fig.~\ref{fig:target_relation}).

\subsection{Defining Places in a System}
\label{sec:target}
\upup

In order to define a place in a system, or a \I{target}, we first need to evaluate the
landscape of all available different targets. Examples for targets are: Sets of CPU cores,
which can be used to solve NUMA related problems; Different memory areas, such
as scratch pads, used to access high bandwidth or other special purpose memory;
Accelerator devices, such as GPUs, which can be used to offload compute
intensive tasks; Remote Processes in a distributed application; Other types of special
purpose hardware, like FPGAs; etc.

Since all the examples for different targets given above have distinct use cases
in terms of their ability to execute different tasks (data parallelism, code with many
control structures) and different properties such as
different address spaces or mechanisms to allocate memory as well as executing code,
it becomes obvious that the definition of a target should be in the form of an
concept that doesn't define the behavior directly, but rather is a handle to an
opaque implementation defined place in the system.
This does not require any additional constraints or specification
for the concept itself, since an implementation is directly operating on
target properties specific to a in place memory and to execute work. By not having those
artificially imposed limitations, we allow for maximal flexibility by defining
the required customization points wherever appropriate.

For supporting the richer functionality to dispatch memory placement and
work execution, the implementation is based on the interfaces described in
Sec.~\ref{sec:allocator} and Sec.~\ref{sec:executor}.

\subsection{Controlling Memory Placement}
\label{sec:allocator}
\upup

After creating a concept for targets (see Sec.~\ref{sec:target}), we discuss the
actual placement of memory. Placement of memory in general
needs to first and foremost handle the allocation of data, but should also cover the transparent
migration (movement) of data to different places of a given target family. For memory allocation
purposes, we leverage the already widely used concept of the \lstinline!Allocator!.

\begin{lstlisting}[
    caption={Extensions to \lstinline!std::allocator_traits! to support
    efficient memory allocation and construction operations for targets as
    described in Sec.~\ref{sec:target}.},
%     \fixme{why is the target() member static? I thought allocators always have an associated target instance...}},
% This is the allocator_trait, the target function needs to be static of course.
    label={lst:allocator}
]
template <typename Allocator>
struct allocator_traits
  : std::allocator_traits<Allocator>
{
  typedef unspecified reference;
  typedef unspecified const_reference;

  typedef unspecified access_target;
  typedef typename access_target::target_type target_type;

  static target_type target(Allocator const& alloc);

  template <typename ...Ts>
  static void bulk_construct(Allocator& alloc, pointer p,
    size_type count, Ts &&... vs);

  static void bulk_destroy(Allocator& alloc, pointer p,
    size_type count) noexcept;
};
\end{lstlisting}

Allocators are already widely used within the C++ standard library (for example
with containers or smart pointers) with the main purpose of encapsulating memory
allocation. This allows for great reuse of the defined concepts in already
existing code and serves our purpose of hiding memory allocation on opaque targets
perfectly. For the sake of making memory allocations efficient on various targets, such
as for discrete GPU memory or remote processes, we introduced backwards compatibly
extensions. List.~\ref{lst:allocator} outlines the traits class which supports
our extensions, the remaining interface follows \lstinline!std::allocator_traits!.
The extensions are optional, and fall back to the requirements for C++ standard
allocators. The extensions introduced, serve the purpose to perform bulk
construction and destruction of C++ objects. This is necessary to either avoid
overheads of offloading the constructor or destructor code or to support
first-touch policies (as used for ccNUMA architectures) efficiently.

The topic of transparent data migration is not covered within this paper and
does not fall within the realm of memory allocation. Another mechanism would
need to be created with appropriate customization points to support different
target use cases. One example within the HPX runtime system is the migration
of objects between different localities (where a locality is a HPX specific
target).

\subsection{Controlling Execution Locality}
\label{sec:executor}
\upup

The previous sections described the mechanisms to define targets (Sec.~\ref{sec:target})
and memory allocation (Sec.~\ref{sec:allocator}). The missing piece, execution
of work close to targets, is based on the \lstinline!Executor! concept. Executors
are an abstraction which define where, how, and when work should be executed, in a
possibly architecture specific way (see also~\cite{cxx14_n4406}).

\begin{lstlisting}[
    caption={\lstinline!std::executor_traits! to support efficient
    execution on targets as described in Sec.~\ref{sec:target}},
    label={lst:executor}
]
template <typename Executor>
struct executor_traits
{
  typedef Executor executor_type;

  template <typename T>
  struct future { typedef unspecified type; };

  template <typename Executor_, typename F, typename ... Ts>
  static void apply_execute(Executor_ && exec, F && f,
    Ts &&... ts);

  template <typename Executor_, typename F, typename ... Ts>
  static auto async_execute(Executor_ && exec, F && f,
    Ts &&... ts);

  template <typename Executor_, typename F, typename ... Ts>
  static auto execute(Executor_ && exec, F && f,
    Ts &&...ts);

  template <typename Executor_, typename F, typename Shape,
      typename ... Ts>
  static auto
  bulk_async_execute(Executor_ && exec, F && f,
    Shape const& shape, Ts &&... ts);

  template <typename Executor_, typename F, typename Shape,
      typename ... Ts>
  static auto
  bulk_execute(Executor_ && exec, F && f,
    Shape const& shape, Ts &&... ts);
};
\end{lstlisting}

Executors follow the same principle as \lstinline!Allocators! in such that they
are accessible through the trait \lstinline!executor_traits!
(see List.~\ref{lst:executor}), in a similar fashion to \lstinline!allocator_traits!.
It is important to note that an implementation for a given executor is not required
to implement all functions as outlined but the traits are able to infer missing
implementations. The only mandatory function an executor needs to be
implement is \lstinline!async_execute!.
The remaining facilities are, if not provided otherwise, automatically deduced
from that.
However, it is important
to note that architectures like GPGPUs benefit tremendously by implementing the
bulk execution features.

Specific executor types are then specialized, architecture dependent
implementations of the executor concept which use this architecture dependent
knowledge to provide the target specific mechanisms necessary to launch
asynchronous tasks.
We introduce a selection of special purpose executors in Sec.~\ref{sec:implementations}.


\subsection{Parallel Algorithms and Distributed Data Structures}
\label{sec:supporting_data_structures}
\upup

Now that we have all the necessary ingredients to co-locate work and data, we are
going to make it usable by providing a specialized implementation of a vector.
This vector is exposing the same high level interface as \lstinline!std::vector<T>!.
This data structure encapsulates an array of elements of the same type
and enables accessing the stored data element-wise, through iterators, and using
other supporting facilities like resizing data, giving the user an abstraction
over contiguous data using the API as described in Sec.~\ref{sec:allocator}.

The exposed iterators can be used directly with the parallel algorithms~\cite{hpx_espm2}
already existing in HPX. Additionally, HPX's parallel algorithms
allow us to pass executors (see Sec.~\ref{sec:executor}) which will in turn execute
the algorithm on the designated resources. By using compatible targets for both, the
executor and the allocator, the co-location of tasks and data is guaranteed.

List.~\ref{lst:hello_world} is providing an example that transforms the string
"hello world" to all uppercase. Note that this example is omitting actual targets
and specific allocators/executors which will be introduced in
Sec.~\ref{sec:implementations}.

\begin{lstlisting}[
    caption={Hello world example using the introduced concepts  \lstinline!Target!,  \lstinline!Allocator!
    and  \lstinline!Executor!},
    label={lst:hello_world}
]
auto target = ...;

target_allocator<char> alloc(target);
vector<char, target_allocator<char>> s(
  {'h', 'e', 'l', 'l', 'o', 'w', 'o', 'r', 'l', 'd'},
  alloc);

target_executor exec(target);
transform(par.on(exec), s.begin(), s.end(),
    [](char c){ return to_upper(c); });
\end{lstlisting}

\section{Implementation}
\label{sec:implementations}
\upup

This Section describes specific implementations for the concepts we defined in
Sec.~\ref{sec:concepts} to demonstrate the feasibility of our claims. As a proof of
concept we implemented special allocators and executors to support NUMA
architectures as well as an allocator and various executors for CUDA devices.

\subsection{Support for NUMA aware programming}
\label{sec:numa}
\upup


Contemporary compute nodes nowadays usually consist of two or more sockets.
Within this architecture, co-locating work and data is an
important ingredient to leverage the full memory bandwidth
within the whole system to avoid NUMA related bottlenecks and to reduce
cross-NUMA-domain (cross-socket) memory accesses.

\begin{figure}
    \begin{center}
    \includegraphics[width=0.99\linewidth]{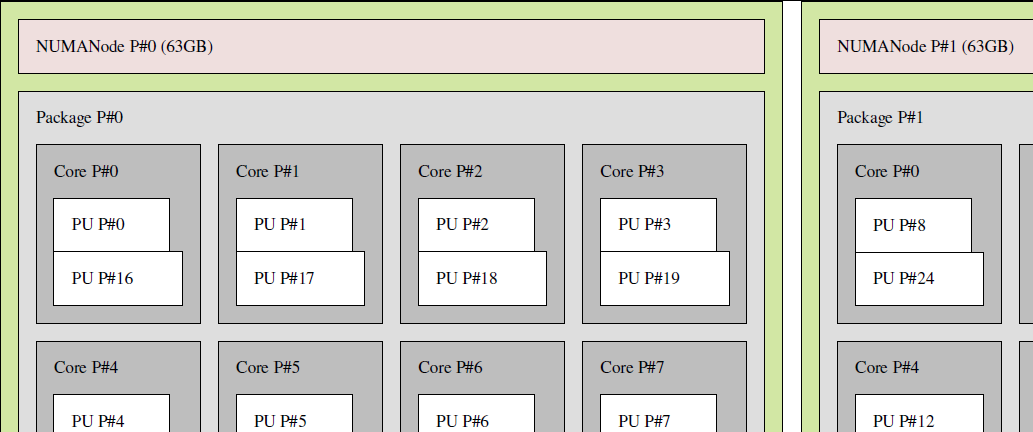}
    \end{center}
\upup
\upup
    \caption{Graphical output from \lstinline!hwloc-ls! showing a two-socket Ivy
    Bridge system. The Figure shows the different identifiers for the different
    processing units and groups them in their respective NUMA domain.\upup
    }
    \label{fig:targets_cpu}
\end{figure}

For this purpose of identifying cores, the target (see Sec.~\ref{sec:target}) is a
numeric identifier for a specific CPU (see Fig.~\ref{fig:targets_cpu}). To
identify the various cores in the system we use hwloc~\cite{broquedis:inria-00429889}. This allows
us to use bit-masks to define multiple CPUs as one single target, making up a
convenient way to build targets for whole NUMA domains. For convenience, two
functions are provided: \lstinline!get_targets()! which returns a vector
containing all processing units found in the system and \lstinline!get_numa_domains()!
which is returning a vector with targets, where each element in that vector
represents the cpuset (bit-mask) identifying the respective NUMA domain. The targets
returned from those functions can then easily be transformed as necssary, for instance
to construct finer grained sets.

After having the targets defined to support our CPU based architecture we need
a target specific allocator to support a vector of separate targets. As a proof
of concept we chose to implement a block allocation scheme by dividing
the number of bytes to be allocated evenly across the passed elements in the
target vector.

The same scheme as described above is used to implement the executor. This ensures
that work that is to be executed on data using the block allocator is co-located with
the data. The cpusets in the associated targets are used as an affinity mask to
pin the executed tasks to the cores accordingly.

\subsection{Leveraging CUDA based GPGPUs}
\label{sec:cuda}
\upup

Since more and more supercomputers are equipped with NVIDIA based GPGPUs as
accelerators, this section will cover a prototypical solution based on CUDA.

Within the CUDA based implementation, the choice for how to define a target is
determined by the underlying programming model. The devices are represented by numerical identifiers and
in order to support asynchronous operations such as kernel invocations and memory
copies, CUDA streams are used. That means a CUDA target is implemented as a wrapper for an
integer representing the device and a CUDA stream attached to that device.

For memory placement, an \lstinline!Allocator! (see
List.~\ref{lst:allocator}) are specialized to allocate memory on the
given device and the (bulk) construct/destruct functions offload directly to the
GPU. In terms of transparent memory access via references to a given object we
introduce a special proxy object that allows to hide the implementation specific
details on how to read and write memory and as such seamlessly supports the
interfaces described in Sec.~\ref{sec:supporting_data_structures}. For copying
data between host and device, we extended the parallel copy algorithm to provide a
internal specialization that is able to directly call the respective CUDA memcpy functions
for maximum efficiency.

The executor support (see Sec.~\ref{sec:executor}) is exploiting the dual
compilation mode of CUDA's nvcc compiler and is therefore able to execute any
callable that is marked with the CUDA specific \lstinline!__device__! attribute.
This gives great flexibility since code that is supposed to be offloaded needs to
be, in theory, only marked with the \lstinline!__device__! (in addition to
\lstinline!__host__!) attribute and can be used immediately with the new executor.

The executor itself implements all functions as outlined in
List.~\ref{lst:executor}. For one, the implemented bulk execution
facility ensures best-possible performance for executing kernels on an array of data-elements.
Secondly, we implemented both, the synchronous and asynchronous versions of the
executor interfaces as the synchronous versions can be implemented more efficiently
than they could be generated by the traits class.
The asynchronous versions additionally need to attach a callback to the stream in order
to notify the returned future about the completion of the operation on the GPU.

In practice however, this way of programming leads to unforeseen problems during
compilation since not all features of C++ are supported on devices and the need
to mark up every function does not scale well especially for third party code.
Newer compiler technologies such as hcc are more promising in this regard and a
truly single source solution without additional mark up can be implemented there
in the near future.

\section{Results}
\label{sec:results}
\upup

For the sake of giving a first evaluation of our abstractions defined in the
previous sections, we are using the STREAM Benchmark~\cite{McCalpin1995}. As a
proof of concept, the presented APIs have been implemented with the HPX parallel
runtime system, and have been ported to the parallel algorithms as defined in the
newest C++ Standard~\cite{cpp17_standard} (see List.~\ref{lst:stream}).

\begin{lstlisting}[
    caption={Generic implementation of the STREAM benchmark using HPX and
    C++ standards conforming parallel algorithms.},
    label={lst:stream}
]
template <typename Executor, typename Vector>
void stream(Executor& e, Vector const& as, Vector const& bs, Vector const& cs)
{
  double scalar = 3.0
  // Copy
  copy(e, as.begin(), as.end(), cs.begin());
  // Scale
  transform(e, cs.begin(), cs.end(), bs.begin(),
      [scalar](double c){ return c * scalar;});
  // Add
  transform(e, as.begin(), as.end(), bs.begin(), cs.begin(),
      [](double a, double b){ return a + b;});
  // Triad
  transform(e, bs.begin(), bs.end(), cs.begin(), as.begin(),
      [scalar](double b, double c){ return b + c*scalar;});
}
\end{lstlisting}

It is important
to note that the benchmark is parameterized on the allocaor used for the arrays (vectors) and the given
executor, which allows to design an portable implementation of the benchmark ensuring
best possible performance across heterogeneous architecturs in plain C++.
For any of the tested architectures, the used \lstinline!Executor! and \lstinline!Vector! is
are using the same target. In our case, we use the NUMA target as defined in Sec.~\ref{sec:numa} and a CUDA target
as described in Sec.~\ref{sec:cuda}.
The test platform we use is a dual socket Intel Xeon CPU E5-2650v2 with 2.60GHz
together with a NVIDIA Tesla K40m GPU. As a reference implementation for the
NUMA based benchmark, we used the original STREAM benchmark~\cite{McCalpin2007},
The GPU version was compared with the CUDA based GPU-STREAM~\cite{Deakin2015}.
For the CPU based experiments, we used two sockets and 6 cores per socket, that
is a total of 12 CPU Cores, which
has been determined to deliver the maximal performance for the benchmark

\begin{figure}
    \centering
    \includegraphics[width=0.70\linewidth]{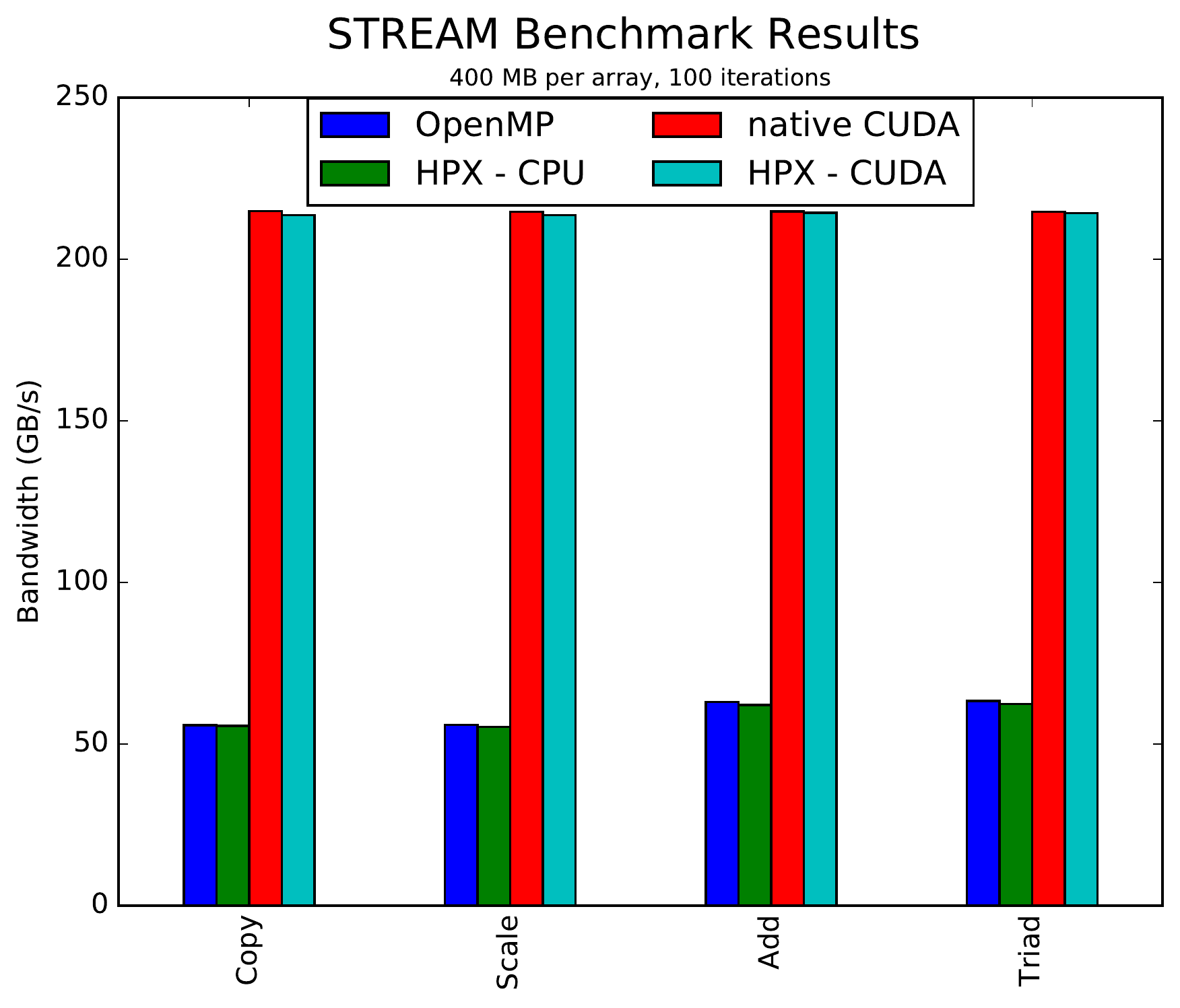}
\upup
\upup
    \caption{Results for STREAM Benchmark on the host and GPU. The figure shows
    the resulting bandwidth achieved with the native implementations and the HPX
    port showed in List.~\ref{lst:stream}. The achieved performance for all tests
    is approximately the same for the respective architectures.
     The benchmark
     on the CPU used 2 NUMA domains with 6 cores each, the GPU version
     ran on a single Tesla K40m.
    }
    \label{fig:stream_results}
\end{figure}

\begin{figure}
    \centering
    \includegraphics[width=0.70\linewidth]{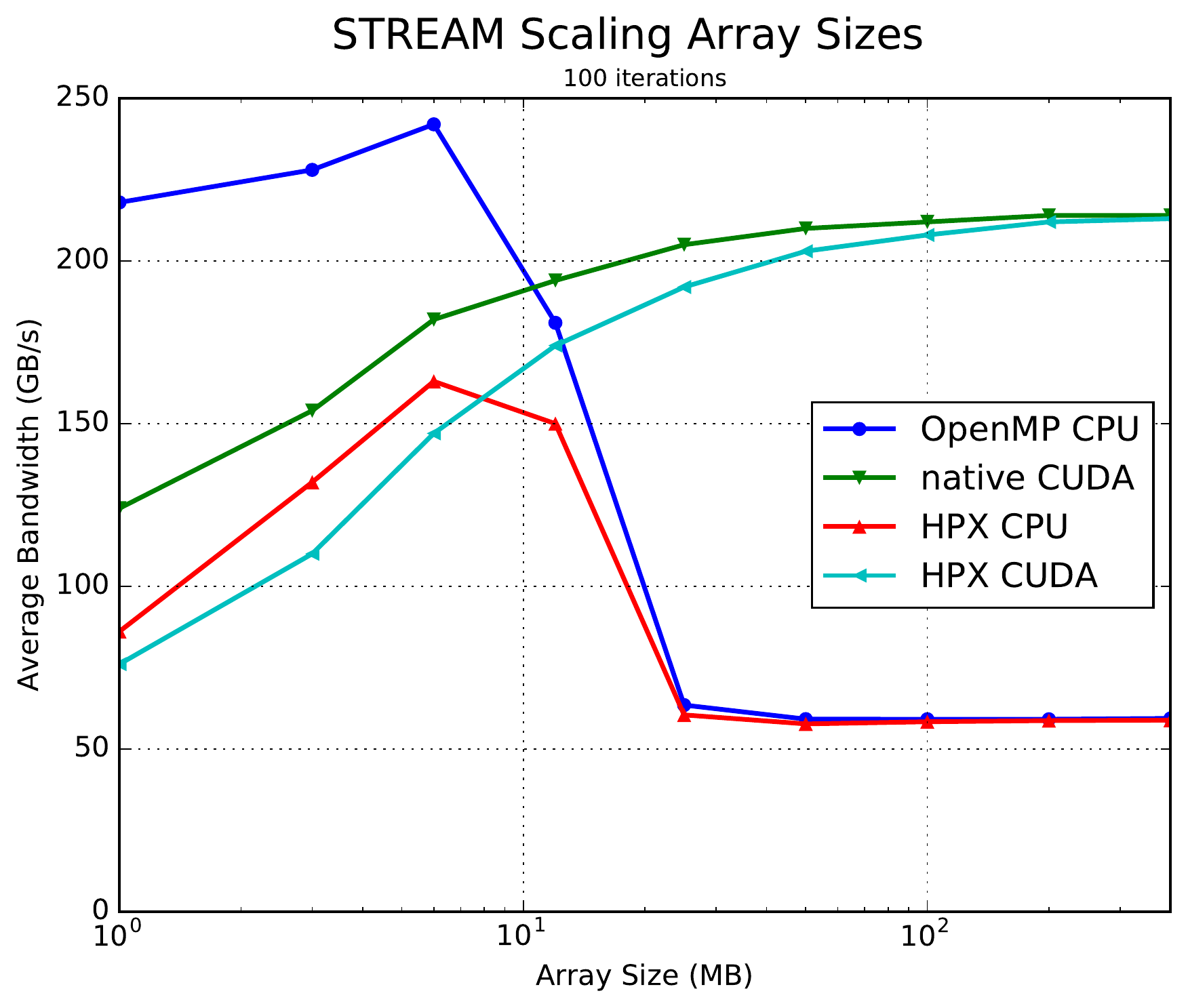}
    \caption{Results for STREAM Benchmark on the host and GPU. This graph shows
    the average performance of the entire benchmark suite with varying input
    sizes, from $10$ to $400$ MB. While the native implementations provide lower overhead for small
    input sizes, all implementations converge to the same maximum.
     The benchmark on the CPU used 2 NUMA domains with 6 cores each, the GPU version
     ran on a single Tesla K40m.}
\upup
\upup
    \label{fig:stream_array_size}
\end{figure}

The results we obtained from running our benchmark show that the utilized memory
bandwidth is essentially equivalent to that achieved by the native benchmarks.
Fig.~\ref{fig:stream_results} is showing results comparing to the respective
reference implementations. What can be seen is that our CPU based implementation
is about $1.1$\% slower than the reference and our CUDA implementation are about $0.4$\% slower.
Fig.~\ref{fig:stream_array_size} is giving an impression on the overheads
involved with the parallel algorithms abstractions. For small array sizes, the
overhead is noticeable within the HPX implementation, however, for reasonable
large extents of the array, the implementations are almost similar again.


\section{Conclusion}
\label{sec:conclusions}
\upup\upup

This paper presented a coherent design and implementation based on the foundation
of the upcoming C++17 Standard and provided extensions to ensure locality of work
and data. We showed that the performance of the introduced higher-level
parallelism framework is does not significantly reduced compared to the performance of
today's prevalent programming environments. The benefit of our presented
solution is to provide a single source, generic, and extensible abstraction for
expressing parallelism, together with no loss in performance.

For future work, we are going to extend the number of targets to include more
support for different memory hierarchies (e.g. Intel Knights Landing High
Bandwitdth Memory) as well as improving the support for GPGPU based solutions
by implementing other back ends such as HCC and SYCL.

\renewcommand{\abstractname}{\ackname}
\begin{abstract}
This work is
supported by the NSF awards 1240655 (STAR), 1447831 (PXFS), and 1339782 (STORM),
and the DoE award DE-SC0008714 (XPRESS) and by the European Union's Horizon 2020
research and innovation program under grant agreement No 671603.
\end{abstract}


\bibliographystyle{splncs03}
\bibliography{bibliography}
%
%
%

\end{document}